\documentstyle[preprint,aps,epsfig]{revtex}
\begin{document}
\tighten
\newcommand{\beq}{\begin{equation}}
\newcommand{\eeq}{\end{equation}}
\newcommand{\bea}{\begin{eqnarray}}
\newcommand{\eea}{\end{eqnarray}}
\newcommand{\ek}{\not{\varepsilon}}
\newcommand{\eek}{\vec{\varepsilon}}
\newcommand{\gp}{ {\cal G\/}_{F} }

\draft
\preprint{BETHLEHEM-Phys-HEP0001019}
\title{Hot nuclear matter in the modified quark-meson coupling model with 
quark-quark correlations} \date{Jan 18, 2000}

\author{
I. Zakout$^{1,2}$ and  H. R. Jaqaman$^{1}$\footnote{hjaqaman@bethlehem.edu}}
\address{$^1$ Department of Physics, Bethlehem University,
Bethlehem, Palestine}
\address{$^2$ Institut f\"ur Theoretische Physik, 
J. W. Goethe Universit\"at, Frankfurt am Main, Germany}  

\maketitle
\begin{abstract}

Short-range quark-quark correlations in hot nuclear matter 
are examined within the modified quark-meson coupling model (MQMC) 
by adding repulsive scalar and vector quark-quark interactions.
Without these correlations, the bag radius increases with 
the baryon density. 
However when  the correlations are introduced
the bag size shrinks as the bags overlap.
Also as the strength of the scalar quark-quark correlation is increased,
the decrease of the effective nucleon mass $M^{*}_N$
with the baryonic density is slowed down and tends to saturate at high densities.
Within this model we study the phase transition from the baryon-meson phase
to the quark-gluon plasma (QGP) phase with the latter  modeled as 
an ideal gas of quarks and gluons inside a bag.
Two models  for the QGP bag parameter are considered.
In one case, the bag  is taken to be medium-independent and 
the phase transition from the hadron phase to QGP is 
found to occur at 5-8 times ordinary nuclear matter density 
for temperatures less than 60 MeV. For lower densities, the transition 
takes place at higher temperature reaching up to 130 MeV at zero density.
In the second case, the QGP bag parameter is considered medium-dependent 
as in the MQMC model for the hadronic phase.
In this case, it is found that the phase transition occurs at much lower densities.

\end{abstract}

\narrowtext
%%%%%%%%%%%%%%%%%%%%%%%
\section{introduction}
The MQMC model has been recently used to study  cold\cite{Jin,Jin2} 
and hot\cite{Jaqaman,Jaqamanb} nuclear matter. 
In this model\cite{Guichon,blunden,Saito94},  nucleons are assumed to be 
nonoverlapping MIT bags interacting through scalar $\sigma$  and vector 
$\omega$ mean fields  
coupled to the quarks themselves. In analogy with 
the nontopological soliton model\cite{soliton}, 
the bag parameter is assumed to decrease 
when the scalar mean field $\sigma$ 
increases\cite{Jin,Jin2}, which makes the bag parameter medium- or 
density-dependent. However, as a result of introducing 
this medium-dependent bag parameter, it is found  
that as the  baryonic density $\rho_B$ increases, 
the nucleon bag radius increases\cite{Jin2,Tsushima}. 
At some value of $\rho_{B}$, the 
bags start to overlap\cite{Saito}.
Since the MQMC model  assumes that the bags do not overlap,
the use of the this model has been limited 
to small and moderate baryonic densities\cite{Jin2}.

One way to extend the model is to include short-range 
quark-quark correlations\cite{Saito} which become important
when the bags overlap at high baryonic densities. 
These correlations are  introduced by adding extra repulsive scalar 
and vector contact  forces between the quarks in the overlapping bags
to reduce the overlapping domain between the nucleons. 
We will follow Saito {\em et al.\/}\cite{Saito} 
and introduce these  correlations in a simple geometrical 
way by defining a critical  rigid-ball nucleonic radius $R_{c}$ which, 
assuming close packing, can be 
related to the baryonic density by  $R_{c}=\left( 1/4\sqrt{2}\rho_{B} 
\right)^{1/3}$.  Hence, for a given nuclear density $\rho_{B}$, 
the nucleon bags are assumed  to overlap only when
the bag radius $R$ is larger than $R_c$.
When the nucleon bags overlap, the quarks in the bags correlate 
with each other by a repulsive potential to reduce the overlapping
effect by shrinking the size of the bags. 
Within this model, we shall study 
the quark-quark correlations for hot nuclear matter 
as well as their effects on the phase transition from 
the hadronic phase to the quark-gluon-plasma
(QGP) phase. The QGP is considered as an ideal gas of 
noninteracting quarks and gluons inside a 
bubble or bag with bag parameter $B$\cite{Serot86}.
It is interesting to study the variation of the phase 
transition with the strength of the quark-quark 
correlation and to examine the possibility that the bag parameter for the
QGP  is also medium-dependent as in the MQMC model.

The outline of the paper is as follows.
In Sect. II, the quark-quark correlations for the overlapping bags
are generalized to the case of hot nuclear matter. In Sect. III we 
introduce a simple model for the QGP. Finally, Sect. III is devoted to our 
results and conclusions.

\section{Quark-quark correlations}

In dense nuclear matter, the nucleons are expected to overlap and  
the quarks in one nucleon correlate 
with the quarks in another. 
This correlation depends basically on how 
much the nucleons overlap with each other.  
The probability $P(R_c/R)$ for two nucleons, each of radius $R$, 
to overlap can be estimated, using a simple geometrical 
approach\cite{Saito,Close}, to be
\begin{eqnarray}
P\left(\frac{R_{c}}{R}\right)=
\left[1-\frac{3}{4}\left(\frac{2R_{c}}{R}\right)+
\frac{1}{16}\left(\frac{2R_{c}}{R}\right)^{3}\right]
\theta\left(\frac{R_{c}}{R}\right)
\theta\left(1-\frac{R_{c}}{R}\right).
\end{eqnarray}

Since the quark-quark correlations are of short range 
it is reasonable to approximate them by a contact interaction
\cite{Saito}.
In the mean-field approximation, the Dirac equation for the quark field 
inside a nucleon bag is given by
\begin{eqnarray}
\left[ i\gamma\cdot\partial-(m_{q}-g^{q}_{\sigma}\sigma
+f^{q}_{s}<\overline{\psi}_{q}\psi_{q}>)
-(g^{q}_{\omega}\omega+f^{q}_{v}<\psi^{\dagger}_{q}\psi_{q}>)\beta
\right]\psi_{q}=0
\end{eqnarray}
where $m_{q}$ is the current  quark mass, 
$f^{q}_{s(v)}$ is the coupling constant for scalar (vector)-type
short-range correlations while 
$<\overline{\psi}_{q}\psi_{q}>$
and
$<\psi^{\dagger}_{q}\psi_{q}>$
are the average values of the quark scalar density 
and quark density\cite{blunden}. The latter, 
following Ref.\cite{Saito},  are approximated 
by $<\overline{\psi}_{q}\psi_{q}>=
\frac{m^{2}_{\sigma}}{g_{\sigma}}\sigma$ and
$<\psi^{\dagger}_{q}\psi_{q}>=3\rho_{B} $.
In the present work, as suggested by Ref.\cite{Saito}, 
the correlation potentials 
are taken  as 
\begin{eqnarray}
f^{q}_{s}<\overline{\psi}_{q}\psi_{q}>=
\alpha P\left(R_{c}/R\right) \sigma,
\end{eqnarray}
and
\begin{eqnarray}
f^{q}_{v}<\psi^{\dagger}_{q}\psi_{q}>=
\beta P\left(R_{c}/R\right) \rho_{B},
\end{eqnarray}
where  $\alpha$ and $\beta$ are parameters used to 
control the strengths of the scalar and vector quark-quark correlations.
Note that as defined here $\alpha$ is a dimensionless parameter, while 
$\beta$ has the dimensions of  $1/(\mbox{Energy})^2$ . 
The coupling constants $g^{q}_{\sigma}$ 
and $g^{q}_{\omega}$ for the scalar and vector mean fields are determined
by reproducing the properties of normal nuclear matter.

The single-particle quark and antiquark energies in units of
$R^{-1}$ are given as
\begin{eqnarray}
\epsilon^{n\kappa}_\pm=\Omega^{n\kappa}\pm 
\left[
g_{\omega}^{q}\omega +f_{v}^{q}<\psi^{\dagger}_{q}\psi_{q}>
\right] R,
\label{EPN} %(2.2)
\end{eqnarray}
where
\begin{eqnarray}
\Omega^{n\kappa}=\sqrt{ x^{2}_{n\kappa} + R^{2}{m^{*}}^{2}_{q} }
\label{Omegnk} %(2.3)
\end{eqnarray}
and 
$m^{*}_{q}=m^{0}_{q}-g^{q}_{\sigma}\sigma
+f^{q}_{s}<\overline{\psi}_{q}\psi_{q}>$ 
is the effective quark mass.
The boundary condition at the bag surface is given by
\begin{eqnarray}
i\gamma\cdot \hat{n} \psi_{q}^{n\kappa}=\psi_{q}^{n\kappa},
\label{roots} % (2.4)
\end{eqnarray}
which determines the quark momentum $x_{n\kappa}$ in the state
characterized by specific values of $n$ and $\kappa$.
%%%%%%%%%
The quark chemical potential $\mu_{q}$, assuming that there are
three quarks in the nucleon bag, is determined through
\begin{eqnarray}
n_{q}&=&3 \nonumber \\
     &=&3\sum_{n\kappa}\left[
 \frac{1}{e^{(\epsilon^{n\kappa}_{+}/R-\mu_{q})/T}+1}
-\frac{1}{e^{(\epsilon^{n\kappa}_{-}/R+\mu_{q})/T}+1}
\right].
\label{nq} % (2.5)
\end{eqnarray}
The total energy from the quarks and antiquarks is
\begin{eqnarray}
E_{\mbox{tot}}=3\sum_{n\kappa}
\frac{\Omega^{n\kappa}}{R}\left[
 \frac{1}{e^{(\epsilon^{n\kappa}_{+}/R-\mu_{q})/T}+1}
+\frac{1}{e^{(\epsilon^{n\kappa}_{-}/R+\mu_{q})/T}+1}
\right].
\label{Etot} % (2.6)
\end{eqnarray}
The bag energy is given by 
\begin{eqnarray}
E_{\mbox{bag}}=E_{\mbox{tot}}
-\frac{Z}{R}+\frac{4\pi}{3}R^{3}B(\sigma).
\label{Ebag} % (2.7)
\end{eqnarray}
where $B(\sigma)$ is the bag parameter.
The medium effects are taken into account for the bag parameter
\cite{Jin}
\begin{eqnarray}
B=B_{0}\exp\left(-\frac{4g^{B}_{\sigma}\sigma}{M_{N}}\right)
\label{BB0} % (2.8)
\end{eqnarray}
where $B_0$ corresponds to a free nucleon
and $g^{B}_{\sigma}$ is an additional parameter.       
The spurious center-of-mass momentum of the bag is subtracted
to obtain the effective nucleon mass
\begin{eqnarray}
M^{*}_{N}=\sqrt{E^{2}_{\mbox{bag}}-<p^{2}_{\mbox{cm}}>}
\label{MNSTAR} % (2.9)
\end{eqnarray}
where
\begin{eqnarray}
<p^{2}_{\mbox{cm}}>=\frac{<x^{2}>}{R^{2}}
\label{PCM} % (2.10)
\end{eqnarray}
and
\begin{eqnarray}
<x^{2}>=3\sum_{n\kappa} x^{2}_{n\kappa}
\left[
 \frac{1}{e^{(\epsilon^{n\kappa}_{+}/R-\mu_{q})/T}+1}
+\frac{1}{e^{(\epsilon^{n\kappa}_{-}/R+\mu_{q})/T}+1}
\right].
\label{x2} % (2.11)
\end{eqnarray}
%%%%%%%%%%%%%%%%%%%%%%%%%%%%%%%%%%%%%%%%%%%%%%%%%%%
The bag radius $R$ is obtained by minimizing the effective   
nucleon mass with respect to the bag radius 
\begin{eqnarray}
\frac{\partial M^{*}_{N}}{\partial R}=0.
\label{MNR} % (2.12)
\end{eqnarray}
The pressure is given by\cite{Jaqaman,Jaqamanb}
\begin{eqnarray}
P=
\frac{1}{3}\frac{\gamma}{(2\pi)^{3}}\int d^{3} k   
\frac{k^{2}}{\epsilon^{*}}(f_{B}+\overline{f}_{B})
+\frac{1}{2}m^{2}_{\omega}\omega^{2}
-\frac{1}{2}m^{2}_{\sigma}\sigma^{2},
\label{pressure} % (2.20)
\end{eqnarray}
where $\gamma=4$ is the spin-isospin degeneracy factor and $f_{B}$ 
and $\overline{f}_{B}$ are the Fermi-Dirac distribution
functions for the nucleons and antinucleons
\begin{eqnarray}
f_{B}=\frac{1}{e^{(\epsilon^{*}-\mu^{*}_{B})/T}+1} \nonumber\\
\overline{f}_{B}=\frac{1}{e^{(\epsilon^{*}+\mu^{*}_{B})/T}+1},
\label{fB} % (2.14)
\end{eqnarray}
with $\epsilon^{*}=\sqrt{ k^{2}+{M^{*}_{N}}^{2} }$ and
$\mu^{*}_{B}=\mu_B-
3\left[g^{q}_{\omega}\omega+f^{q}_{v}<\psi^{\dagger}_{q}\psi_{q}>\right]$
being the nucleonic effective energy 
and effective chemical potential, respectively.
The chemical potential $\mu_B$ for a given density $\rho_{B}$ 
is determined self-consistently by the subsidiary 
constraint\cite{Jaqaman,Jaqamanb}
\begin{eqnarray}
\rho_{B}=\frac{\gamma}{(2\pi)^{3}}\int d^{3}k(f_{B}-\overline{f}_{B})
\label{rhoB} % (2.16)
\end{eqnarray} 
with 
\begin{eqnarray}
\omega=\frac{g_{\omega}}{m^{2}_{\omega}}\rho_{B}.
\label{omegrho} % (2.17)
\end{eqnarray}
The scalar mean field $\sigma$ is determined 
through maximizing the pressure  
$\frac{\partial P}{\partial \sigma}=0$ which yields the self-consistency 
condition (SCC) for the $\sigma$ field. 
Since the scalar type correlation does not
directly involve the $\sigma$ field,
the SCC is not formally modified by it and is therefore identical to that 
found in our earlier work\cite{Jaqaman} .
The correlations do however affect the $\sigma$ field indirectly through
the quark wave functions.
%%%%%%%%%%%%%%%%%%%%%%%%%%%%%%%%%%%%%%%%%%%%%
%%%%%%%%%%%%%%%%%%%%%%%%%%%%%%%%%%%%%%%%%%%%%
\section{The Quark-Gluon Plasma phase}

In the QGP phase we assume that we have only 
$u$ and $d$ quarks confined inside a bag with bag parameter B.
This parameter can be interpreted as the energy per unit volume 
needed to create a bubble or bag 
in which the noninteracting quarks and gluons 
are confined.
The total baryonic density is given by\cite{Serot86}
\begin{eqnarray}
\rho_B=\frac{1}{3} \frac{\gamma_Q}{(2\pi)^3}
\int d^{3} k\left[n_k(T)-\overline{n}_k(T)\right].
\label{QGP1}
\end{eqnarray}
The quark and antiquark distribution functions are given by
\begin{eqnarray}
n_{k}=\frac{1}{\exp[(k-\frac{1}{3}\mu_B)/T]+1}, \nonumber\\
\overline{n}_{k}=\frac{1}{\exp[(k+\frac{1}{3}\mu_B)/T]+1},
\end{eqnarray}
respectively,
where $\mu_B$ is the baryon chemical potential, and we have 
assumed that the quarks have a  baryon number of $1/3$.
At finite $\rho_B$, Eq.(\ref{QGP1}) is inverted 
to find the baryon chemical potential $\mu_B$.
The pressure of the quark-gluon plasma is given by\cite{Serot86}
\begin{eqnarray}
P=-B+\frac{1}{3}\frac{\gamma_{G}}{2}\frac{T^4\pi^2}{15}
+\frac{1}{3}\frac{\gamma_{Q}}{(2\pi)^{3}}
\int d^{3}k k\left[n_k(T)+\overline{n}_{k}(T)\right]
\end{eqnarray}
where $k=|\vec{k}|$ and $\gamma_{Q}=12$ for quarks
and $\gamma_{G}=16$ for gluons.
The thermodynamic conditions for phase equilibrium
between the baryon-meson phase and the QGP phase are satisfied by
assuming mechanical, chemical and  thermal equilibrium between 
the two phases namely, 
$P_{\mbox{MQMC}}=P_{\mbox{QGP}}$ and 
${\mu_B}_{\mbox{MQMC}}={\mu_B}_{\mbox{QGP}}$
for a given $T$. These conditions determine the transition 
line between the hadron-meson phase and the QGP phase.

We consider two cases. In the first case, the bag parameter 
in the QGP phase is considered as medium-independent 
and has the same value as that for a free nucleon. This case 
may be appropriate for most experimental situations 
aiming at producing the QGP in small chunks of hot nuclear 
matter produced  in heavy ion collisions. In the second case, the bag 
parameter in the QGP is assumed to be medium-dependent, and 
is taken to be equal to the bag parameter of the 
nucleon in the MQMC model at the same density. This corresponds to the 
idealized case of producing the QGP in a bubble in infinite hot nuclear matter 
and may approximately apply to the production of the 
QGP in central collisions between very massive nuclei.

%%%%%%%%%%%%%%%%%%%%%%%%%%%%%%%%%%%%%%%%%%%%%
%%%%%%%%%%%%%%%%%%%%%%%%%%%%%%%%%%%%%%%%%%%%%
\section{Results and Discussions}

We have studied nuclear matter at finite temperature using the MQMC model 
which takes the medium-dependence of the bag parameter of the nucleon into account. 
We choose a direct coupling of the bag parameter to the
scalar mean field $\sigma$ as given in Eq.\ref{BB0}. 
The bag parameters are those given in Ref.\cite{Jin,Jaqaman,Jaqamanb}
where they are chosen to reproduce the free 
nucleon mass $M_{N}$ at its experimental value of 939 MeV 
and a bag radius $R_{0}=0.60$ fm.
For $g^{q}_{\sigma}=1$, the values of the vector meson coupling 
and the parameter $g^{B}_{\sigma}$, as fitted from the normal
saturation properties of nuclear matter\cite{Jin}, are given as 
$g^{2}_{\omega}/4\pi=5.24$ and $(g^{B}_{\sigma})^{2}/4\pi=3.69$.
The  current quark mass $m_{q}$ is taken equal to zero.
For the short-range quark-quark correlation strengths we use values 
comparable to those in Ref.\cite{Saito} who, 
in the present notation, use 
$\alpha \approx 9 $ and $\beta \approx 34\mbox{GeV}^{-2}$.
The latter value  of $\beta$ is needed to reproduce 
the empirical value of the energy per nucleon for 
symmetric nuclear matter in the high density region 
$\rho_B/\rho_0=2.5-4$, where $\rho_0=0.17 \mbox{fm}^{-3}$ 
is normal nuclear density. 
We carried out calculations for the cases of  $\alpha=$ 0, 5, 7 
and 9 as well as $\beta=$ 0, 30 and 60 GeV$^{-2}$. 
The scalar quark-quark correlations affect  
the nucleon's size and mass while 
the vector quark-quark correlations determine
the pressure and energy density of nuclear matter\cite{Saito}.

Fig. 1 displays isotherms of the effective 
mass $M^{*}_N$ vs the baryonic density
$\rho_B$ for various strengths of the scalar 
quark-quark correlations. As already mentioned, 
the vector correlations do not have an effect on the mass.
For low values of $\alpha$, $M^{*}_N$ has the usual trend
of decreasing with $\rho_B$.
However, as $\alpha$ increases
$M^{*}_N$ tends to saturate and,  
for still larger values of $\alpha$, the effective mass
even starts to increase  slightly at high density, especially 
at the lower temperatures. This novel feature is quite 
interesting as it is questionable that the monotonic 
decrease of $M^{*}_N$ with density can continue 
unchecked for higher and higher densities.

Fig. 2 displays isotherms of $R/R_0$ vs the baryon density for several
values of $\alpha$.
Without correlations, {\em i.e.} $\alpha=0$, $R/R_0$ increases 
monotonically with
$\rho_B$. However, when the correlations are introduced,
$R/R_0$ starts to decrease, rather abruptly, when the bags start to overlap.
This abruptness is due to the simple geometrical 
way in which the correlations are introduced.
As $\alpha$ is increased further, $R/R_0$ decreases more steeply 
at high densities. The repulsive nature 
of the quark-quark correlations shrinks the bag size.
%%%%%%%%%%%%%%%%%%%%%%%%

Fig. 3 displays the transition line between the baryon-meson phase
and the QGP phase in the $(T,\mu_{B})$ plane, 
while Fig. 4 displays it in the $(T,\rho_B)$ plane.
This phase transition line is determined by equalizing  the pressure $p$ and chemical
potential $\mu_{B}$ in both phases 
for a given temperature $T$.
We have considered two cases for the bag parameter of the QGP bubble.
In  case I, the bag parameter
is taken to be medium-independent and fixed at its free-space value.
As the density $\rho_B$ is increased, the nucleon bags start to overlap
and the transition line becomes sensitive to the quark-quark correlations for  
 $\mu_B$ larger than about 950 MeV, corresponding 
to densities larger than about 2.5 times normal nuclear 
matter density.
For  temperatures $ T< 60$ MeV, the 
phase transition takes place at rather large 
densities and large chemical potentials.  The quark-quark 
correlations are found to move the transition line to 
lower chemical potentials and lower densities. For cold nuclear 
matter the correlations can reduce $\mu_{B}$ 
from 1850 to 1450 MeV. The corresponding change 
in the  $(T,\rho_B)$ plane is more dramatic as 
can be seen by inspecting Fig. 4. The phase 
transition, at low temperatures, takes place at  densities
as high as 8$\rho_0$ without correlations, but this value 
is reduced to about 5$\rho_0$ for the strongest correlations 
considered.
In  case II,  also shown in Figs. 3 and 4,  we have
used a medium-dependent bag parameter
for the QGP bubble. This bag parameter  is identical to the 
bag parameter $B(\sigma)$ used  for the nucleonic 
bags in the hadronic phase as given in Eq.\ref{BB0}. 
This medium-dependence is appropriate for the production 
of a QGP bubble in infinite nuclear and may be appropriate 
for its production in the heart of  the participant region in central 
collisions between very massive nuclei. In this case it is found that  
the phase transition from the baryon-meson
phase to the QGP phase takes place at much lower 
densities so that  the nucleons do not overlap and the
quark-quark correlations do not play a role in determining the transition line.
 The  transition temperature falls rapidly with density and the phase  
transition at low  temperatures occurs at a comparatively low chemical potential
$\mu_{B}=950$ MeV and a  correspondingly low density 
$\rho_B / \rho_0=1.35$ . This density is obviously too low for the 
production of the QGP in heavy ion collisions and is strictly
appropriate only  for infinite nuclear matter as it does 
not include any finite size effects. It does however hint 
at the sizable reduction in the compression needed to 
produce the QGP in collisions between very heavy systems.

In conclusion we have investigated the effect of 
short-range quark-quark correlations on the properties 
of hot nuclear matter and the phase transition to the QGP. 
We have found that these correlations cure the 
problem usually encountered  in the MQMC model of 
a very large nucleonic bag radius. They also lead to the saturation
of the effective mass at high densities. 
Moreover, these correlations affect the properties 
of the phase transition at low temperatures for the case 
of a medium-independent bag parameter for the QGP 
bubble in vacuum (case I). Such a situation arises in experimental 
situations attempting to produce the QGP in small finite 
hot nuclear systems. In such a case, the present results 
indicate that the phase transition occurs at very high 
densities 5-8 times normal nuclear matter density. 
The only exceptions occur at very high temperatures 
greater than 100 MeV, in which case the transition 
occurs at arbitrarily small densities. In 
case II we have used a medium-dependent bag parameter 
for the QGP bubble, and the phase transition 
is found to occur at much lower densities than in Case I. 
The phase transition occurs before the nucleons 
overlap and so the quark-quark correlations do not 
play a role in determining the transition line.
This case, strictly speaking,  corresponds to producing  
a QGP bubble in infinite nuclear matter, but 
may  be approximately approached in central collisions 
involving two very heavy nuclei. The comparatively low compressions 
required for the phase transition in such collisions would thus 
offer the best chance of producing the QGP.

%%%%%%%%%%%%%%%%%%%%%
%%%%%%%%%%%%%%%%%%%%%
\acknowledgments
Financial support by  the Deutsche Forschungsgemeinschaft through the  
grant GR 243/51-1 is gratefully acknowledged.

%%%%%
\clearpage

% Fig. 1
\begin{figure}
\centerline{\mbox{\epsfig{file=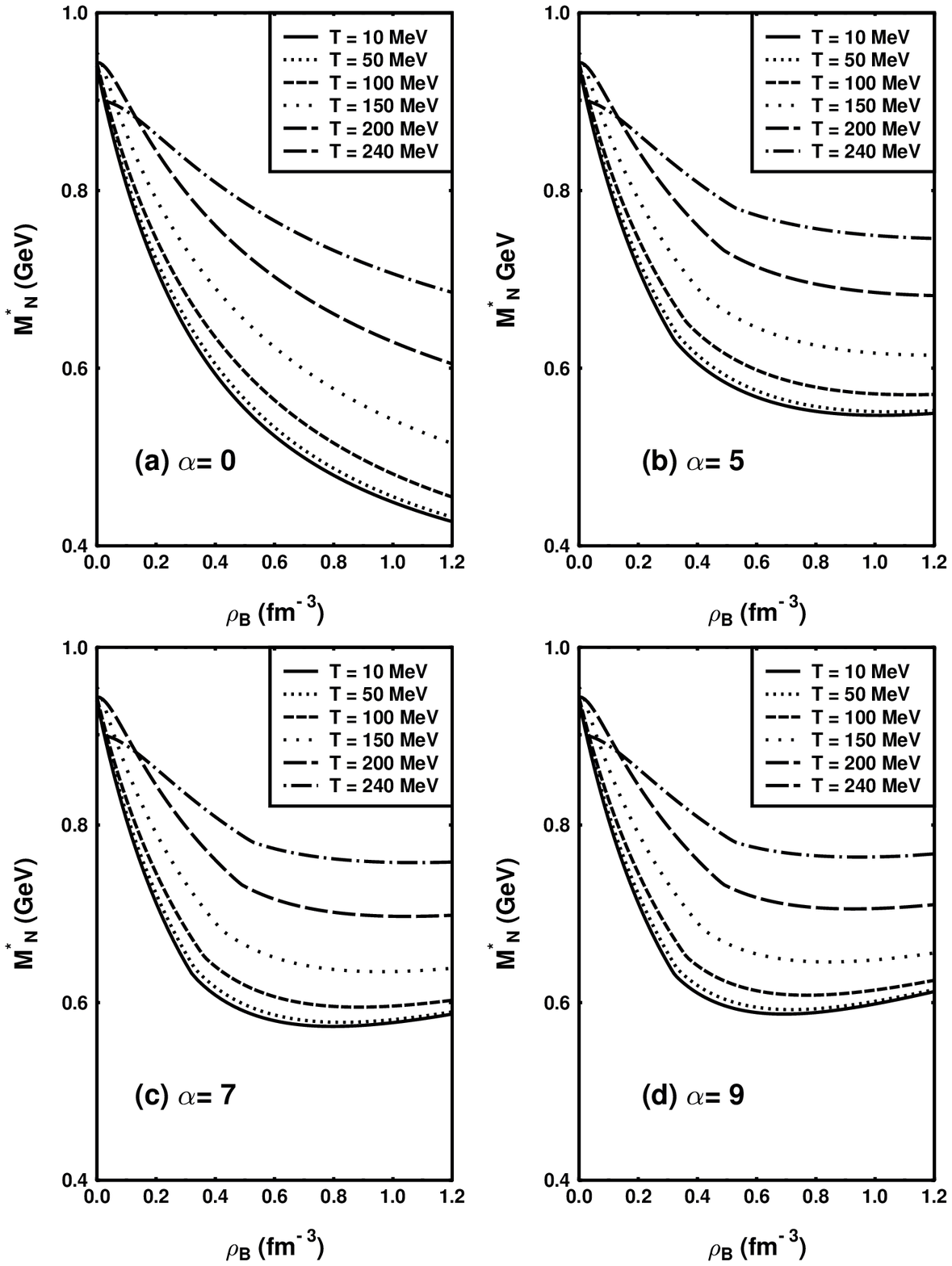,width=0.8\linewidth}}}
\vspace{1truein}
\caption{
Isotherms of the effective nucleon mass $M^{*}_{N}$ 
as a function of the baryonic
density $\rho_{B}$  for different strengths
of the scalar quark-quark correlations.}
\end{figure}

% Fig. 2
\begin{figure}
\centerline{\mbox{\epsfig{file=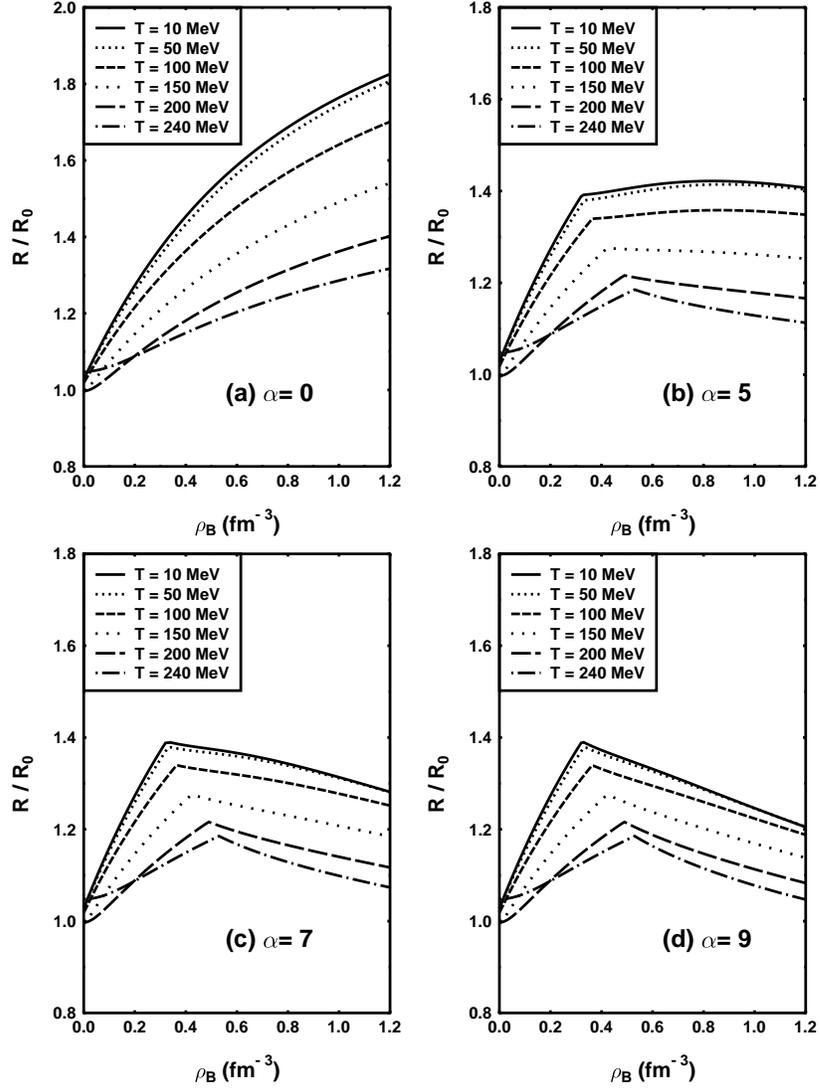,width=0.8\linewidth}}}
\vspace{1truein}
\caption{
Isotherms of  $\frac{R}{R_{0}}$ 
vs the baryonic density $\rho_{B}$ 
for different strengths of the scalar quark-quark correlations.}   
\end{figure}

% Fig. 3 
\begin{figure}
\centerline{\mbox{\epsfig{file=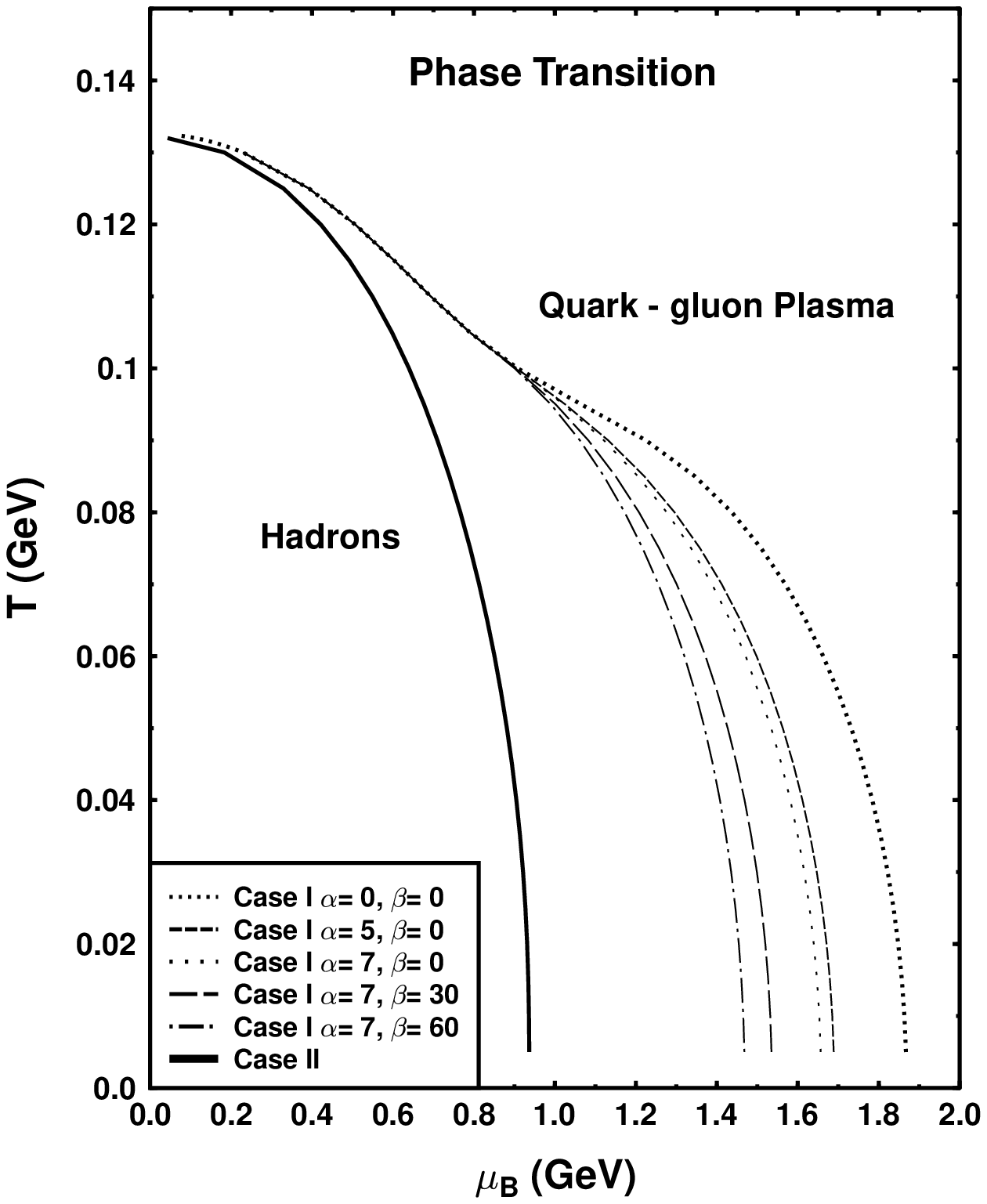,width=0.8\linewidth}}}
\vspace{1truein}
\caption{  
The phase transition line, in the $(T,\mu_{B})$ plane, 
between the  hadronic phase and the  QGP
phase for different strengths of the scalar 
and vector quark-quark correlations.}   
\end{figure}

% Fig. 4 
\begin{figure}
\centerline{\mbox{\epsfig{file=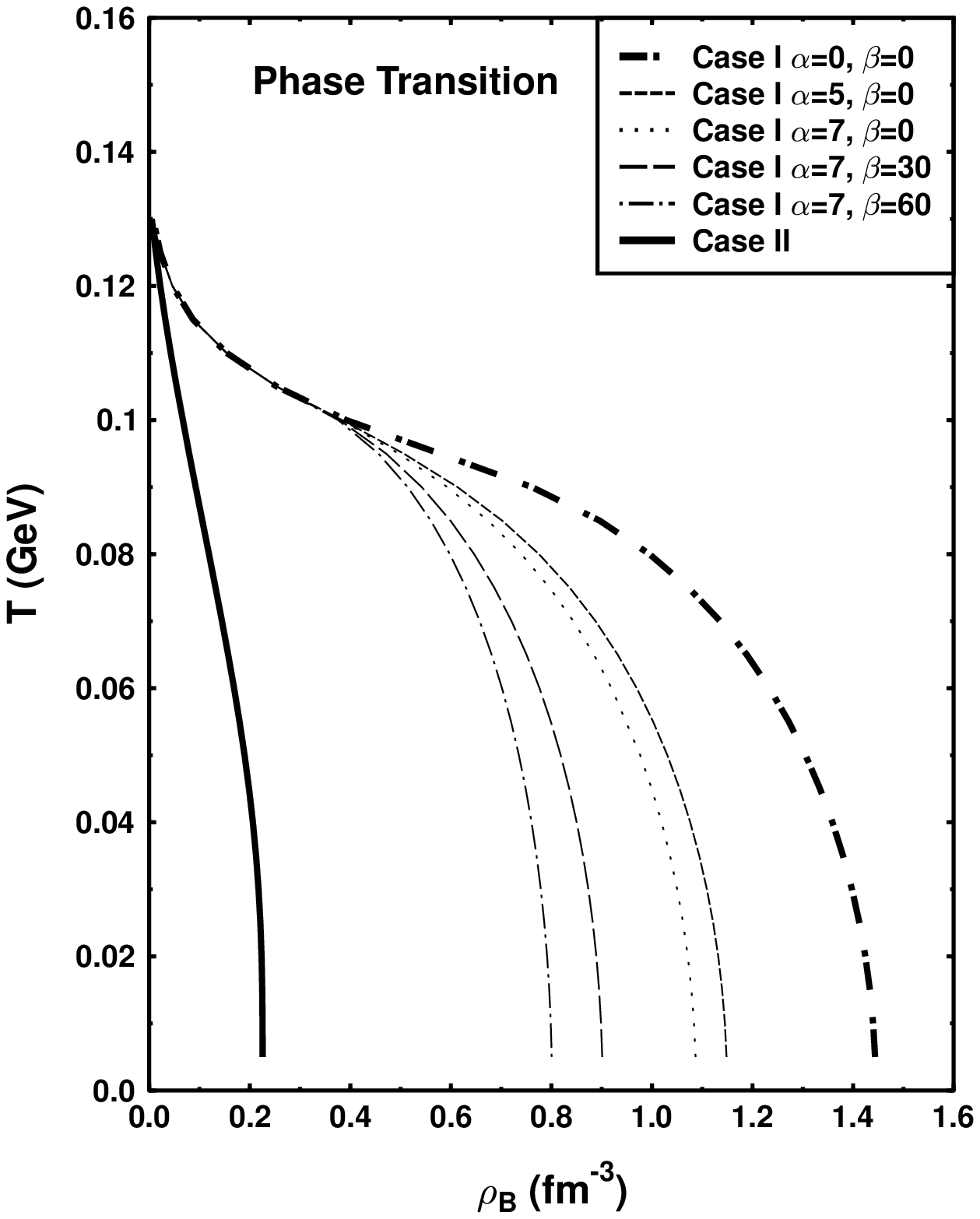,width=0.8\linewidth}}}
\vspace{1truein}
\caption{     
The onset phase transition line, in the $(T,\rho_{B})$ plane, between 
the hadronic phase and the QGP phase for different strengths  
of the scalar and vector quark-quark correlations.}    
\end{figure}


\begin{references}
%%%%%%%%%%%%%%%%%%%%%
%% modified quark meson coupling, bag radius increase
\bibitem{Jin}
X. Jin and B. K. Jennings, Phys. Rev. C {\bf 54}, 1427 (1996);
Phys. Lett. B {\bf 374}, 13 (1996).
%% Problem in the bag radius where it is increased
\bibitem{Jin2}
X. Jin and B. K. Jennings, Phys. Rev. C {\bf 55}, 1567 (1997);
H. M\"uller and B. K. Jennings, Nucl. Phys.  {\bf A 626}, 966 (1997);
H. M\"uller, Phys. Rev. C {\bf 57}, 1974 (1998).   
%%%%%%%%%%%%%%%%%%%%%
% hot nuclear matter
\bibitem{Jaqaman}
I. Zakout and H. R. Jaqaman, Phys. Rev. C {\bf 59}, 962 (1999).
% Hot nuclear matter
\bibitem{Jaqamanb}
I. Zakout and H. R. Jaqaman, Phys. Rev. C {\bf 59}, 968 (1999).
% QMC formalism, QMC model proposed
\bibitem{Guichon}
P. A. M. Guichon, Phys. Lett. B {\bf 200}, 235 (1988).
%%%%%%%%%%%%%%%%%%%%%
%  <psi|psi>, <psi^{\dagger}|psi>
\bibitem{blunden}
P. G. Blunden, G. A. Miller, Phys. Rev. C {\bf 54}, 359 (1996).
% QMC applied for nuclear matter
\bibitem{Saito94}
K. Saito and A. W. Thomas, Phys. Lett. B {\bf 327}, 9 (1994).
% The soliton model
\bibitem{soliton}
M. K. Banerjee, Phys. Rev. C {\bf 45}, 1357 (1992);
V. K. Mishra, Phys. Rev. C {\bf 46}, 1143 (1992);
E. Naar and M. C.  Birse, J. Phys. G. {\bf 19}, 555 (1993).
% Problem in the bag radius where it is increased
\bibitem{Tsushima}
D. H. Lu, K. Tsushima, A. G. Williams, A. W. Thomas and K. Saito,
Nucl. Phys. {\bf A 634}, 443 (1998).
% Quark-Quark correlations
\bibitem{Saito}
K. Saito, K. Tsushima and A. W. Thomas, nucl-th/9901084.
%%%%%%%%%%%%%%%%%%%%%%%%%
%\bibitem{Walecka75} % QHD method
%J. D. Walecka, Ann. of Phys. {\bf 83}, 491 (1974);
%Phys. Lett. B {\bf 59}, 109 (1975).
% review mean field model and nuclear application
\bibitem{Serot86}
B. D. Serot and J. D. Walecka, Adv. Nucl. Phys. {\bf 16}, 1 (1986).
%%%%%%%%%%%%%%%%%
% The overlapping probability
\bibitem{Close}
F. E. Close, R. L. Jaffe, R. G. Roberts and G. G. Ross,
Phys. Rev. D {\bf 31}, 1004 (1985).
\end{references}
\end{document}